\documentclass{elsart}

\usepackage{natbib}

\usepackage{graphicx}
\usepackage{amssymb}

\begin{document}

\begin{frontmatter}



\title{A Heavenly Example of Scale Free Networks and Self-Organized Criticality}


\author{M. Paczuski\corauthref{cor}}
\corauth[cor]{Corresponding author}
\ead{maya@ic.ac.uk}
\author{and D. Hughes}

\address{Mathematical Physics, Imperial College London, London,
UK  SW7 2BZ}

\begin{abstract}
% Text of abstract
The sun provides an explosive, heavenly example of self-organized
criticality.  Sudden bursts of intense radiation emanate from rapid
rearrangements of the magnetic field network in the corona.
Avalanches are triggered by loops of  flux that reconnect or
snap into lower energy configurations when they are overly
stressed. Our recent analysis of observational data reveals that the
 loops (links) and footpoints (nodes), where they attach on
the photosphere, embody a scale free network.  The statistics of
the avalanches and of the  network structure are unified through a simple
dynamical model where the avalanches and network co-generate each
other into a complex, critical state.  This particular example
points toward a general dynamical mechanism for self-generation of
complex networks.

\end{abstract}

\begin{keyword}
% maximum of four
networks \sep self organized criticality \sep solar flares

\PACS  96.60.Rd \sep  64.60.Lx
% See http://www.aip.org/pacs/pacs.html
% or  http://www.elsevier.com/locate/pacs

\end{keyword}

\end{frontmatter}

% main text
\section{Introduction}
\label{} Two generative themes to grasp complexity in its variety of
manifestations confront us.  One refers to a general dynamical
principle, or how complexity is made.  Another refers to ways of
understanding inter-relations between parts of a complex system.  We
will leave aside the problem of defining complexity
\citep{grassberger}, which to date has  been posed in terms of
algorithms or information, rather than plausible dynamical laws.

The first theme we discuss is self-organized criticality
\citep{btw,HowNatureWorks}, which posits that the dynamics of slowly, driven
dissipative systems impels them toward a robust dynamical attractor that
accommodates avalanches of all sizes.  These avalanches, in turn, imprint long
range correlations throughout the structure in which they occur. The
physical system and the avalanches in it cogenerate each other through a long,
historically contingent process into a complex entity. One cannot speak of one,
the system, without the other, the avalanches.

Often, we are mislead by the intermittent occurrence of the avalanches
into thinking of the two as being separate.  Although most of the
time, the system, such as the network of faults in the crust of the
earth, appears static, one cannot forget that a dynamics, in this case
seismic processes or earthquakes, must account for the current
appearance of the network of faults as being a particular snapshot in
geological time over which the crust of the earth, and network of
faults has changed greatly.  (Remembering this fact, one becomes
skeptical of definitions of complexity which forget about the dynamics
of the actual physical process that generates the ensemble of observed states.)

As the second theme, modern network theory has emerged as a powerful
framework for describing many disparate systems ranging from the
topology of the world wide web, to biological regulatory networks
\citep{Barabasi02}, or physical systems such as the coronal magnetic
field \citep{Hughes:longpaper}.  By using nodes as the basic objects,
a great simplification in the description of the system can be
obtained by only linking strongly interacting or correlated nodes.
For example, since $N$ nodes can have at most $N(N-1)/2$ links between
them, pruning insignificant links can produce drastic simplifications,
especially if the final number of links is of order $N$. Then, the
network can be considered as the essential backbone of the
inter-relationships between different parts of the system. In this
way, networks may distill the main qualities, or most significant
features, from the  messy and hard-to-describe systems we
generally refer to as complex.

Empirically, it has been observed that many networks
are neither totally random nor completely ordered like a regular lattice.  They
lie in between, exhibiting strong heterogeneity from node to node.
In fact, some networks are scale-free, in the
sense that the number of links attached to any node is distributed as
a power law, up to a cutoff determined by the total number of links in the network.

Various models of complex or scale-free networks have been proposed.
In the preferential attachment model of Barabasi and Albert, a network
grows by adding links to nodes with a probability proportional to the
number of already existing links at that node \citep{Barabasi99}.  A
common feature of this and other models is that the process generating
the network assures that a system wide network always exists.
Furthermore, the number of nodes and links in the network are assumed
to always grow.

However, one can imagine a different scenario where the number of
nodes and links in the network are statistically
stationary. Furthermore, and perhaps even more to the point, one can
consider a situation {\it where the system wide network only emerges
after a transient self-organizing process}.  Before reaching the
stationary state, no system wide network exists; it arises as a result
of a long, historically contingent process.  Such networks can be said
to be self-organized networks, and in this case one can ask if self-organized
criticality (SOC)
provides a mechanism for the emergence of scale-free networks, out of
a collection of uncorrelated random networks.  This could happen just
like a sandpile, with avalanches of all sizes,
can be made out of a collection of uncorrelated sand grains by slowly
adding them together without disturbing the system too much, for
instance, by shaking the sandpile violently in a continuous fashion.
We now give a specific example of a SOC mechanism for scale-free  networks.
We describe a naturally occuring physical system, namely the sun.

A complex interwoven network of magnetic fields threads the surface of the sun.
Magnetic energy stored in the coronal network builds up due to turbulent plasma
forces until stresses are suddenly released by a rapid reconfiguration of the
magnetic field \citep{mhd}. This releases energy that is converted to radiation and observed as a solar
flare. The probability distribution of flare energies is  a featureless
power law that spans more than eight orders of magnitude
\citep{aschwanden:TRACE}.

Lu and Hamilton first proposed that the corona is in a self-organized critical
state \citep{lh}, with avalanches of all sizes \citep{btw}. However, SOC
systems often show scale free behavior not only for their event statistics, but
also for emergent spatial and temporal structures \citep{btw}, as seen both in
physical systems (e.g. \citep{frette:realrice}) and numerical models (e.g.
\citep{hughes:nsdm}).

In fact, like flares, concentrations of magnetic flux on the photosphere
(which is the visible surface of the sun) also
exist on a wide variety of scales. The strongest concentrations are sunspots,
which occur in active regions that may contain more than $10^{22}$Mx. The
smallest resolvable concentrations above the current resolution scale of
$\approx 10^{16}$Mx are fragments. For many years, solar physicists have
believed that at each scale a unique physical process is responsible for the
dynamics and generation of magnetic concentrations, e.g. a ``large scale dynamo'' versus a ``surface dynamo'' etc.

\begin{figure}
\centering
\includegraphics[width=350pt]{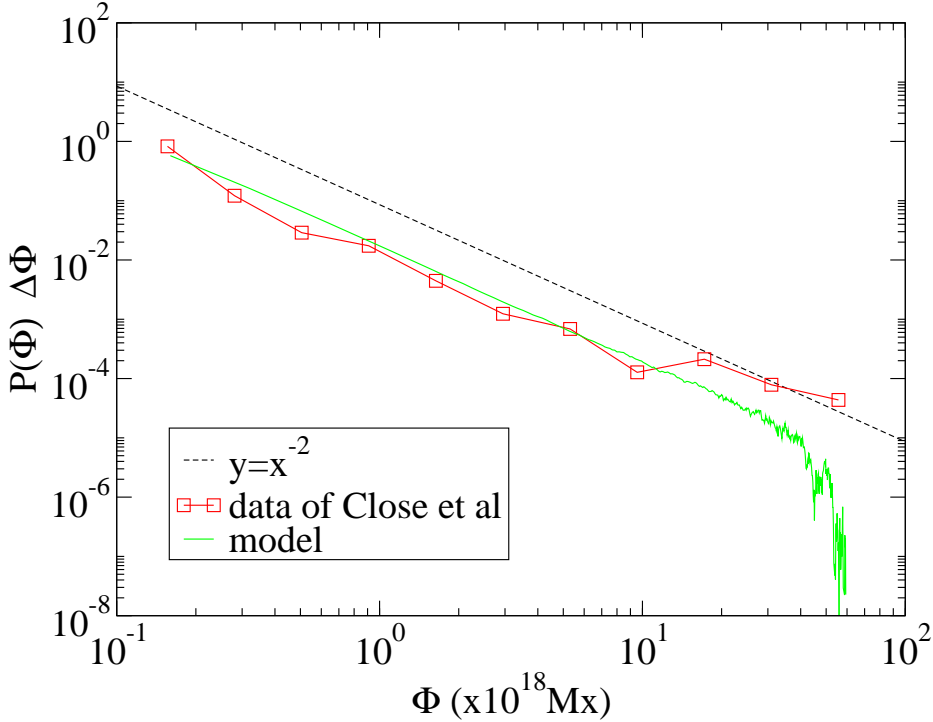}
\caption{The degree distribution of the magnetic network. The normalized number
of magnetic concentrations in bins of size $\Delta \Phi = 1.55\times 10^{17}$
Mx obtained by reanalyzing the measurement data originally shown in Figure 5 of
Ref. \citep{close}. The model data shown represents the probability
distribution, $P(k_{foot})$, for number of loops, $k_{foot}$, connected to a
footpoint. This has been rescaled so that one loop, $k_{foot}=1$, equals the
minimum threshold of flux, $1.55 \times 10^{17}$ Mx }\label{concentrations}
\end{figure}

Motivated by results from numerical simulations of a model, which is described later,
we recently reanalyzed several
previously published data sets reporting the distribution of concentration sizes
\citep{close}. We just plotted the distribution using a double logarithmic scale
instead of a log-linear one, as had been done previously. As shown in Fig.
\ref{concentrations}, we found that the distribution of magnetic flux
concentration sizes is scale free over more than two decades, corresponding to
the range of this particular measurement \citep{Hughes:longpaper}.

Of course the basic equations that describe the evolution of the magnetic field are
continuous. However, the corona is a highly turbulent system and the dynamics
of its magnetic field is intermittent. The field itself is
concentrated into narrow regions and does not spread out uniformly.
Observations of the solar corona reinforce this view, showing loops of hot
plasma, emanating from magnetic concentrations, which are thought to trace out
the paths of the magnetic flux tubes. So, although the fundamental equations of
plasma physics are continuous and smooth, what we actually observe are well
defined, discrete entities such as magnetic concentrations and flux tubes. We
decided therefore to treat the coronal magnetic field as made up of discrete
interacting loops \cite{modelprl}.

\begin{figure}
\centering
\includegraphics[width=350pt]{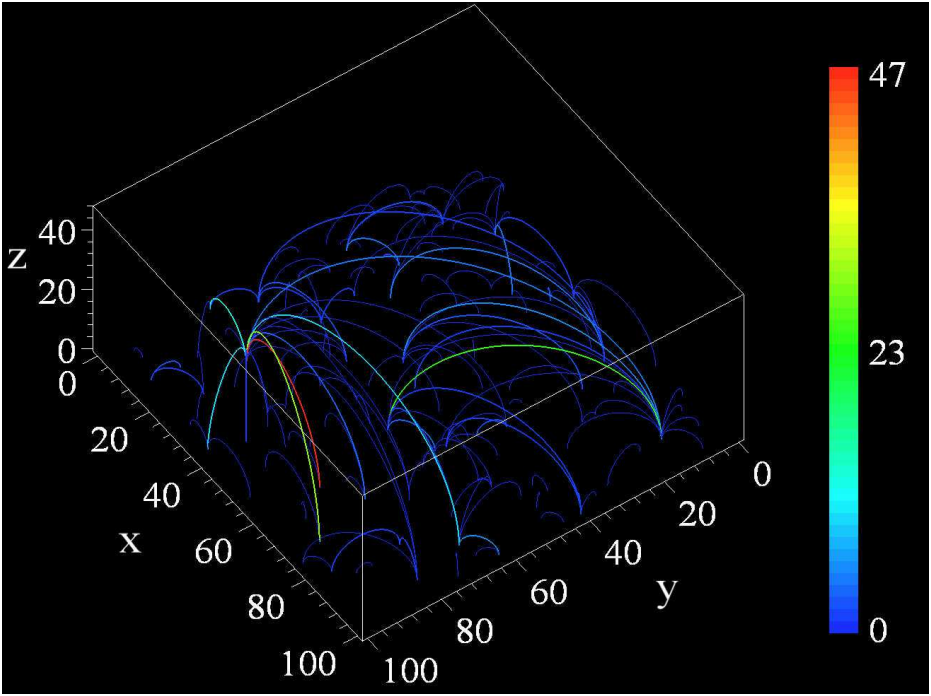}
\caption{Snapshot of loops in the steady state of the model. Footpoints lie in
the $(xy)$ plane and are linked by loops. The loops are colored to indicate
the relative strength of the connection, as shown by the scale to the right.
Note that there is a large range in the number of loops emanating from
different footpoints, as well as a wide range of connection
strengths.}\label{snapshot}
\end{figure}

The fundamental entity in the model is a directed loop that traces the midline
of a flux tube, and is anchored to a flat surface at two opposite polarity
footpoints.  A footpoint locates the center of a magnetic concentration.
A collection of these loops and their footpoints gives a distilled
representation of the coronal magnetic field structure. Our network model is
able to describe fields that are very complicated or interwoven.  A snapshot of
a configuration in the steady-state is shown in Fig. \ref{snapshot}. The number
of loops connecting any pair of footpoints is indicated by a color-coding. It
is evident that both the number of loops attached to a footpoint and the
number  of loops connecting any pair of footpoints vary over a broad range.

Loops injected at small length scales are stretched and shrunk as their
footpoints diffuse over the surface. Nearby footpoints of the same polarity
coalesce, to form magnetic fragments, which can themselves coalesce to form
ever larger concentrations of flux, such as  sunspots. Conversely,
adjacent, opposite polarity footpoints cancel. Each loop carries a single unit
of flux, and the magnetic field strength at a footpoint is given by the number
of loops attached to it. The number of loops that share a given pair of
footpoints measures the strength of the link.

Loops can reconnect when they collide in three dimensional space. The
flux emerging from the positive footpoint of one of the reconnecting
loops is then no longer constrained to end up at the other footpoint
of the same loop, but may instead go to the negative footpoint of the
other loop. This re-wiring process allows footpoints to exchange
partners and reshapes the network, but it maintains the degree of each
footpoint.

If rewiring occurs, it may happen that one or both loops need to cross
some other loop. Thus a single reconnection between a pair of loops
can trigger an avalanche of causally related reconnection
events. Reconnections occur instantaneously compared to the diffusion
of footpoints and injection of loops.

The system therefore evolves by a combination of slow diffusion of
footpoints, occassional injection of new loops and rapid avalanches of
reconnection.  These avalanches punctuate quiet periods where the
configuration of loops only changes gradually and slowly.  Thus, a type of
stick-slip dynamics is observed.

From arbitrary initial conditions, the system eventually reaches a
steady state, where the cancellation processes on average balance the
steady inflow of new loops. In the steady state we find that the system
of loops and footpoints self organize into a scale free network
\citep{Hughes:longpaper}. The number of loops, $k_{foot}$, connected to any
footpoint is distributed as a power law
$$ P(k_{foot})\sim k_{foot}^{-\gamma}\,\, , \qquad \gamma\approx
2\quad .$$ Clearly, each footpoint can be thought of as a node in the
network  and each loop as a link.  Therefore, this distribution
corresponds to the degree distribution of the network, and the
exponent $\gamma$ is the exponent for the degree distribution.

There are a variety of ways to associate an energy with the loops,
depending on their length, and how many loops connect a pair of
footpoints.  Reconnection will shorten loops and therefore liberate
energy from the magnetic field. A cascade of reconnections represents
a flare.  Defining the energy to be only
proportional to the total length of loops, the distribution of flare
energies follows a power law \citep{modelprl}.  However, during the
transient while the system organizes itself, the distribution of flare
energies is not a power law. Thus, our model is a true SOC system.

We find two new quantities to characterize scale free networks.
Numerical simulations of the model show that the strength of the link between
two nodes, which is the number of loops connecting two nodes,
 is distributed as a power law, with an exponent $\alpha\geq\gamma$.
 In addition, the number of unique nodes
linked to any specific node also exhibits scale free behaviour, with a
different exponent, which is greater than $\alpha$.

To sum up, we have demonstrated that SOC is a mechanism for generating
scale free networks. In our model the complex network which
self-generates itself into a state with avalanches of all sizes is
fully scale free. Not only is the degree distribution a power law, but
so is the distribution of link strengths, the number of distinct nodes
linked to any particular node, and the avalanches.  The system wide
network emerges only after a transient period where the avalanches and
the network have cogenerated each other into a critical state.  Before
that happens, no system wide network exists and the avalanches are
also limited by a scale smaller than the system size.

% \appendix

% \section{}
% \label{}

% Bibliographic references with the natbib package:
% Parenthetical: \citep{Bai92} produces (Bailyn 1992).
% Textual: \citet{Bai95} produces Bailyn et al. (1995).
% An affix and part of a reference:
%   \citep[e.g.][Ch. 2]{Bar76}
%   produces (e.g. Barnes et al. 1976, Ch. 2).

\bibliographystyle{elsart-num}
\bibliography{refs}

\end{document}